\documentclass[a4paper]{jpconf}
\usepackage{graphicx,color}
\usepackage{hyperref}
\usepackage{amsmath}
\usepackage{amssymb}
\usepackage{subfig}
\usepackage[square,numbers,sort&compress,comma]{natbib}
\usepackage{cmap}
\hypersetup{colorlinks=true,citecolor=blue}
\begin{document}
\newcommand{\subfigureautorefname}{\figureautorefname}

\newcommand{\ie}{i.e., }
\newcommand{\Schro}{Schr\"o\-din\-ger }
\newcommand{\eg}{e.g.\@ }
\newcommand{\cf}{cf.\@ }
\renewcommand{\equationautorefname}{Eq.}
\renewcommand{\figureautorefname}{Fig.}
\newcommand{\refeq}[1]{\hyperref[#1]{\equationautorefname~(\ref*{#1})}}
\newcommand{\appref}[1]{\hyperref[#1]{appendix~\ref*{#1}}}
\newcommand{\CC}{\ensuremath{\text{C}_{60}}}
\newcommand{\tEWS}{t_{\scriptstyle\textrm{EWS}}}
\newcommand{\tEWSsr}{t_{\scriptstyle\textrm{EWS}}^\textrm{sr}}
\newcommand{\tCLC}{t_{\scriptstyle\textrm{CLC}}}
\newcommand{\tdLC}{t_{\scriptstyle\textrm{dLC}}}
\newcommand{\tS}{t_{\scriptstyle\textrm{S}}}
\newcommand{\tR}{t_{\scriptstyle\textrm{R}}}
\newcommand{\tcc}{t_{\scriptstyle\textrm{cc}}}
\newcommand{\tCLCcl}{t_{\scriptstyle\textrm{CLC}}^{\scriptstyle\textrm{cl.}}}
\newcommand{\tScl}{t_{\scriptstyle\textrm{S}}^{\scriptstyle\textrm{cl.}}}
\newcommand{\tEWScl}{t_{\scriptstyle\textrm{EWS}}^{\scriptstyle\textrm{cl.}}}
\newcommand{\tEWSCcl}{t_{\scriptstyle\textrm{EWS}}^{\scriptstyle\textrm{C, cl.}}}
\newcommand{\tEWSC}{t_{\scriptstyle\textrm{EWS}}^{\scriptstyle\textrm{C}}}
\newcommand{\tCoul}{t_{\scriptstyle\textrm{Coul}}}
\newcommand{\tCoulcl}{t_{\scriptstyle\textrm{Coul}}^{\scriptstyle\textrm{cl.}}}
\newcommand{\ev}{\,\text{eV}}
\newcommand{\eV}{\ev}
\newcommand{\au}{\,\text{a.u.}}
\newcommand{\nm}{\,\text{nm}}
\newcommand{\He}{\text{He}}
\newcommand{\Hede}{\ensuremath{\text{He}^{**}}}
\newcommand{\Hep}{\ensuremath{\text{He}^{+}}}
\newcommand{\Wcm}{\,\text{W}/\text{cm}^2}
\newcommand{\as}{\,\text{as}}
\newcommand{\fs}{\,\text{fs}}
\newcommand{\cvec}[1]{\mathbf{#1}}
\newcommand{\eqcomma}{\,,}
\newcommand{\bra}[1]{\langle#1|}
\newcommand{\ket}[1]{|#1\rangle}
\newcommand{\braket}[2]{\langle#1|#2\rangle}
\newcommand{\dE}{\dd E}
\newcommand{\dk}{\dd k}
\newcommand{\dd}{\mathrm{d}}
\newcommand{\expval}[1]{\langle#1\rangle}
\newcommand{\abs}[1]{\left|#1\right|}
\newcommand{\norm}[1]{\left\|#1\right\|}
\newcommand{\subscr}[1]{_{\scriptstyle\mathrm{#1}}}

\title{Time-resolved photoemission using attosecond streaking}

\author{S Nagele$^{1}$,
R Pazourek$^{1}$,
M Wais$^{1}$,
G Wachter$^{1}$, and
J Burgd\"orfer$^{1}$}

\address{$^1$ Institute for Theoretical Physics, Vienna University of Technology, 1040 Vienna, Austria, EU}

\ead{stefan.nagele@tuwien.ac.at}

\begin{abstract}

We theoretically study time-resolved photoemission in atoms as probed by attosecond streaking. We review recent advances in the study of the photoelectric effect in the time domain and show that the experimentally accessible time shifts can be decomposed into distinct contributions that stem from the field-free photoionization process itself and from probe-field induced corrections. We perform accurate quantum-mechanical as well as classical simulations of attosecond streaking for effective one-electron systems and determine all relevant contributions to the time delay with attosecond precision.
In particular, we investigate the properties and limitations of attosecond streaking for the transition from short-ranged potentials (photodetachment) to long-ranged Coulomb potentials (photoionization). As an example for a more complex system, we study time-resolved photoionization for endohedral fullerenes $A$@$\CC$ and discuss how streaking time shifts are modified due to the interaction of the $\CC$ cage with the probing infrared streaking field.

\end{abstract}

\section{Introduction}

The photoelectric effect, \ie the emission of an electron after the absorption of a photon, is one of the most fundamental processes in the interaction of light with matter. The progress in the creation of ultrashort light pulses during the last decade (see \cite{Agostini04,Corkum07,KliVra2008,KraIva2009,NisSan2009} for a review) has enabled the time-resolved study of photoemission with attosecond precision ($1\as = 10^{-18}$s) \cite{SchFieKar2010,KluDahGis2011}.
One of the most important experimental techniques is attosecond streaking \cite{DreHenKie2001,HenKieSpi2001,ItaQueYud2002,KitMilScr2002,GouUibKie2004,CavMueUph2007,SchFieKar2010} which is a pump-probe scheme that employs a phase-controlled (few-cycle) infrared (IR) field and an extreme ultraviolet (XUV) attosecond pulse to determine the arrival time of the photoelectrons in the continuum. Closely related techniques include RABBIT (``reconstruction of attosecond harmonic beating by 
interference of two-photon transitions'') \cite{VenTaiMaq1996,PauTomBre2001,TomMul2002,KluDahGis2011} and the attosecond clock for (close to) circularly polarized fields \cite{EckSmoSch2008,PfeCirSmo2011}. 
In the following we will review recent developments in the understanding of time delays in photoemission as probed by attosecond streaking
and present results on the transition from short-ranged to long-ranged atomic potentials corresponding to the transition from photodetachment to photoionization.
We also report on the effects of the $\CC$ shell on streaking of photoemission from the central atom of an endohedral complex A@$\CC$.
Atomic units are used throughout the manuscript unless indicated otherwise.

\section{Computational Methods}
 
To simulate streaking experiments we numerically solve the time-dependent \Schro equation for a single-active electron. 
We employ the well-established pseudo-spectral split-operator method \cite{TonChu1997} using a finite-element discrete-variable-representation (FEDVR) \cite{ResMcc2000,SchCol2005} for the discretization of the wave function in coordinate space.
The interaction with the linearly-polarized laser fields is treated in the length form of the dipole approximation. We have carefully tested the numerical convergence of our simulations regarding box size, radial discretization, angular momenta basis, and time-stepping.

\begin{figure}[htbp]
  \centering
  \includegraphics[width=0.6\linewidth]{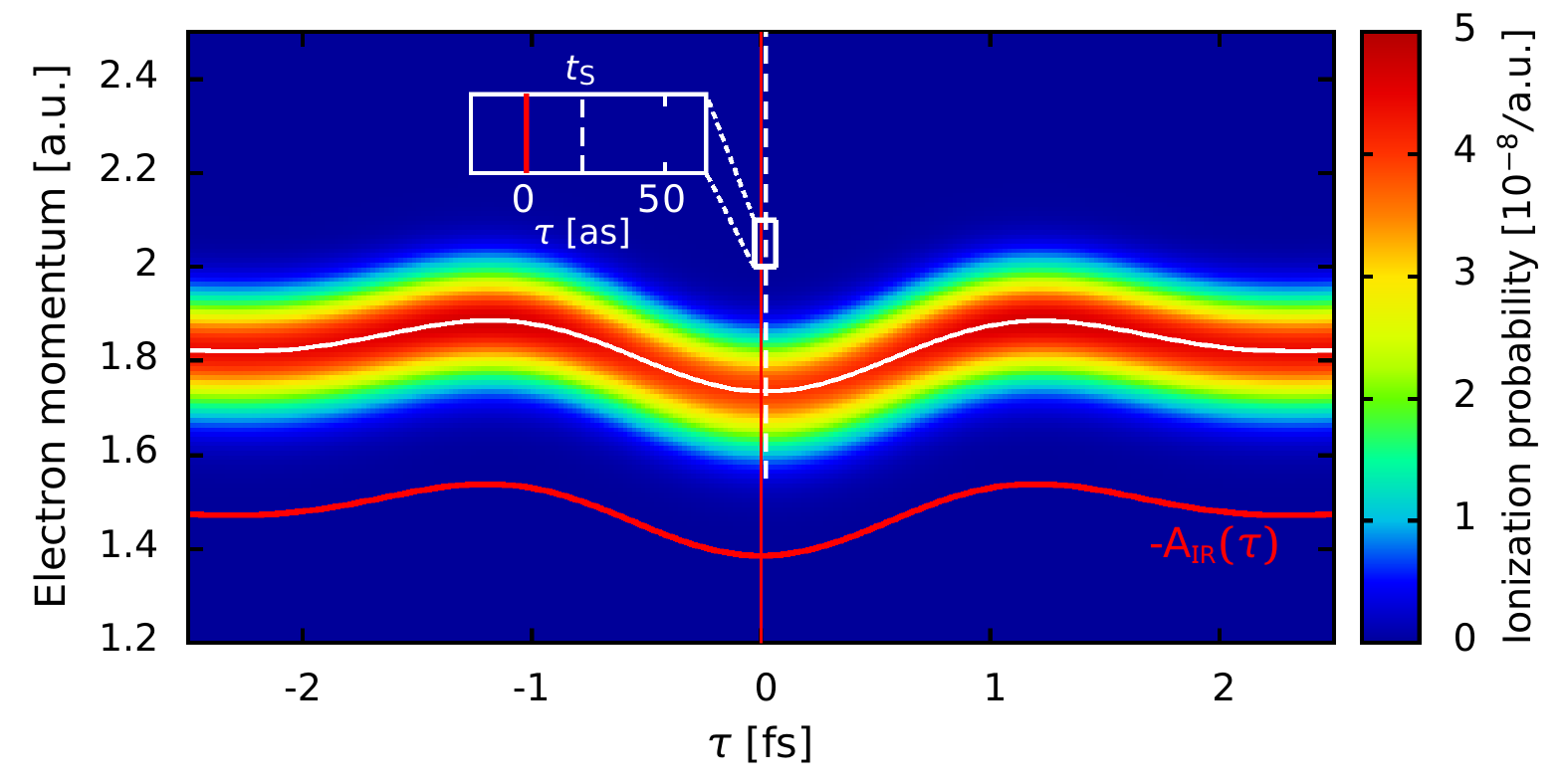} 
  \caption{Streaking spectrogram for an $800\nm$ IR laser field with a duration of 3\,fs and an intensity of $10^{12}$\,W/cm$^2$. The graphs show the final momentum distribution integrated over an opening angle of $10^\circ$ in ``forward'' direction of the laser polarization axis for an $\Hep(1s)$ initial state and an XUV photon energy of 100\,eV (duration $200\as$ FWHM). The solid white line is the first moment of the final electron momentum spectrum which follows the vector potential of the IR laser field (solid red line) (\autoref{eq:streaking_fit}). The small shift $\tS$ between the two curves (see \autoref{eq:streaking_fit}) is revealed in the inset.
  \label{fig:streaking-spectrogram}}
\end{figure}

From the numerical calculations we extract the absolute streaking time shifts $\tS$ by fitting the first moments (or peak positions) of the final momentum distribution $\cvec p_f(\tau)$, which depends on the delay time $\tau$ between the XUV and IR pulses (see \autoref{fig:streaking-spectrogram}), to the IR-modified momentum,
\begin{equation}
\label{eq:streaking_fit}
 \cvec p_f(\tau) = \cvec p_0 - \alpha\cvec A_{\mathrm{IR}}(\tau+\tS) \eqcomma
\end{equation}
using an iterative, nonlinear least squares algorithm, where $\cvec p_0$ is the unperturbed asymptotic momentum of the photoelectron, $\cvec A_{\mathrm{IR}}$ is the vector potential of the IR field, and $\alpha$ is a correction factor for the amplitude of the momentum shift induced by the streaking field. 

If the ionization process and the transition from the bound initial state to the asymptotic momentum $\cvec p_0$ is assumed to happen instantaneously, $\tS$ vanishes and \autoref{eq:streaking_fit} reduces to the standard streaking equation \cite{ConTarSto1997,DreHenKie2001,ItaQueYud2002,KitMilScr2002}.
If the photoelectrons are analyzed in ``forward'' direction (at $\theta=0^\circ$) with respect to the laser polarization axis then $\alpha > 0$, while $\alpha < 0$ for analysis in ``backward'' direction ($\theta=180^\circ$).  The sign convention for $\tS$ implies that \emph{positive} (negative) values correspond to \emph{delayed} (advanced) emission relative to the center of the XUV pulse. 
Typically, averaging the photoelectron spectra over a finite opening angle mainly affects the constant $\alpha$ and has only little effect on $\tS$. 
However, if the photoelectrons have a pronounced anisotropic angular distribution, 
also the streaking delays $\tS$ may become significantly angle-dependent. 

\section{Attosecond streaking of photoemission}

The physical principle underlying attosecond streaking is that for photoemission in the presence of a few-cycle IR field the final momentum (or energy) of the photoelectrons emitted by the XUV pulse is shifted as a function of the delay time between the XUV and IR  pulses (\autoref{eq:streaking_fit}). This delay time $\tau$ can be experimentally controlled with high accuracy leading to a temporal resolution for $\tS$ close to the single-attosecond scale \cite{SchFieKar2010}.
The challenge in interpreting the streaking time shifts obtained lies in disentangling the intrinsic timing of the photoionization process in the absence of the IR probe field, the so-called Eisenbud-Wigner-Smith (EWS) delay \cite{Eis1948, Wig1955, Smi1960}, from additional probe-field induced contributions.

For photoemission the EWS delay is given by the spectral derivative of the dipole matrix element,
\begin{equation}
 \label{eq:EWS_def_matel}
 \tEWS(E,\theta,\varphi) = \frac{\dd}{\dd E} \arg\left[\bra{\psi_f(E,\theta,\varphi)}\cvec d\ket{\psi_i}\right] \, ,
\end{equation}
where $E$ is the energy of the photoelectron and $\cvec d$ is the dipole transition operator between the initial bound state $\psi_i$ and the final scattering state $\psi_f$ describing photoemission into the solid angle ($\theta,\varphi$). 

Alternatively, the EWS time delay can be directly extracted from the motion of the outgoing electronic wave packet \cite{deCNus2002,PazNagBur2013} created by the XUV pulse. After the wave packet has left the scattering region, the time evolution of its radial expectation value $\expval{r}_t$ (or the position of its crest) eventually follows asymptotically $(t \rightarrow \infty)$ the motion of a free particle delayed by the time $\tEWS$ \cite{BreHaa1959},
\begin{equation}
\label{eq:r_t}
 \expval{r}_t= v_g (t - \tEWS) \, ,
\end{equation}
where $v_g$ is the group velocity of the photoelectron wave packet.
The EWS time delay $\tEWS$ can thus be extracted from the intercept of the linear extrapolation of $\expval{r}_t$ with the $t$ axis \cite{PazNagBur2013}, 
\begin{equation}
\label{eq:EWS_rkt}
\tEWS = t - \frac {\expval{r}_t} {v_g} \, .
\end{equation}
However, one key condition of the applicability of Eqs.\ \ref{eq:EWS_def_matel} - \ref{eq:EWS_rkt} is that the binding potential is short-ranged such that the wave packet reaches, indeed, the free-particle motion at large distances \cite{Wig1955,Smi1960}. Note that just as for the streaking delays $\tS$ the sign convention for $\tEWS$ in Eqs.\ \ref{eq:EWS_def_matel} - \ref{eq:EWS_rkt} is such that \emph{positive} (negative) values correspond to \emph{delayed} (advanced) emission.

In general, the measured streaking time shifts $\tS$ do not give direct access to the EWS time $\tEWS$ (Eqs.\ \ref{eq:EWS_def_matel} and \ref{eq:EWS_rkt}).
Several additional contributions have been identified \cite{BagMad2010,BagMad2010Err,ZhaThu2010,ZhaThu2011,NagPazFei2011,PazFeiNag2012,PazNagDob2012,NagPazFei2012},
\begin{equation} \label{eq:delay_relation_full}
  \tS = \tEWS + \tCLC + \tdLC^{(i)} + \tdLC^{(\mathrm{e-e)}} \, ,
\end{equation}
where the Coulomb-laser coupling (CLC) term $\tCLC$ stems from the simultaneous interaction of the Coulomb field and the external laser field on the outgoing electron \cite{ZhaThu2010,NagPazFei2011}, while the dipole-laser coupling terms $\tdLC^{(i)}$ and $\tdLC^{(\mathrm{e-e)}}$ result from the polarization of the atomic initial state or the final ionic state, respectively \cite{BagMad2010,BagMad2010Err,NagPazFei2011,PazFeiNag2012}.
The three additive correction terms to the EWS time have in common that they are all induced by the simultaneous presence of the IR streaking field.
Their relative importance depends on the electronic structure of the residual atomic or molecular complex from which the photoelectron is emitted.

The CLC contribution originates from the interplay between the Coulomb-like long-ranged part of the binding potential and the IR field.
One of its key properties is its independence of the atomic species.
It only depends on the photoelectron energy, the asymptotic ionic charge $Z$, and the wavelength of the IR field and can therefore readily be taken into account for a given experiment or simulation.
Note that an analogous term has also been identified for the complementary, interferometric RABBIT technique \cite{KluDahGis2011, DahGueKlu2012}.
The appearance of the CLC contribution is a direct consequence of the long-range nature of the Coulomb potential where, strictly speaking, EWS delays as given in \autoref{eq:EWS_def_matel} or \autoref{eq:EWS_rkt} are not well-defined (\cf \cite{Wig1955,Smi1960,BolGesGro1983,Mar1981}) due to the logarithmic divergence of the energy and position dependent phase of the Coulomb wave, 
\begin{equation}\label{eq:Coulphase_full}
\phi^\mathrm{Coul} (E,\ell,r) = \sigma_\ell^C (E) + \frac {Z} {k} \ln (2kr) \, ,
\end{equation}
where $k=\sqrt{2E}$ is the asymptotic momentum, $\ell$ the angular quantum number, and 
\begin{equation}\label{eq:Coulphase}
\sigma_\ell^C (E) = \arg \Gamma (1 + \ell - i{Z}/{k}) \, .
\end{equation}
The $r$-independent term \autoref{eq:Coulphase}, often referred to as the ``Coulomb phaseshift'' or ``Coulomb phase'', is the Coulomb analogue to the phase shift in standard scattering theory for short-ranged potentials and its spectral derivative determines the EWS delay for photoionization (\autoref{eq:EWS_def_matel}) which reduces to
\begin{equation}
\label{eq:EWS_def_Coulphase_l}
t_{\mathrm{EWS}}^C (E,\ell) = \frac{\partial}{\partial E} \sigma_\ell^C (E)
\end{equation}
if only a single partial wave in the continuum is accessed.

\begin{figure}[htb]
  \centering
  \subfloat[][]{
  \includegraphics[width=0.5\linewidth]{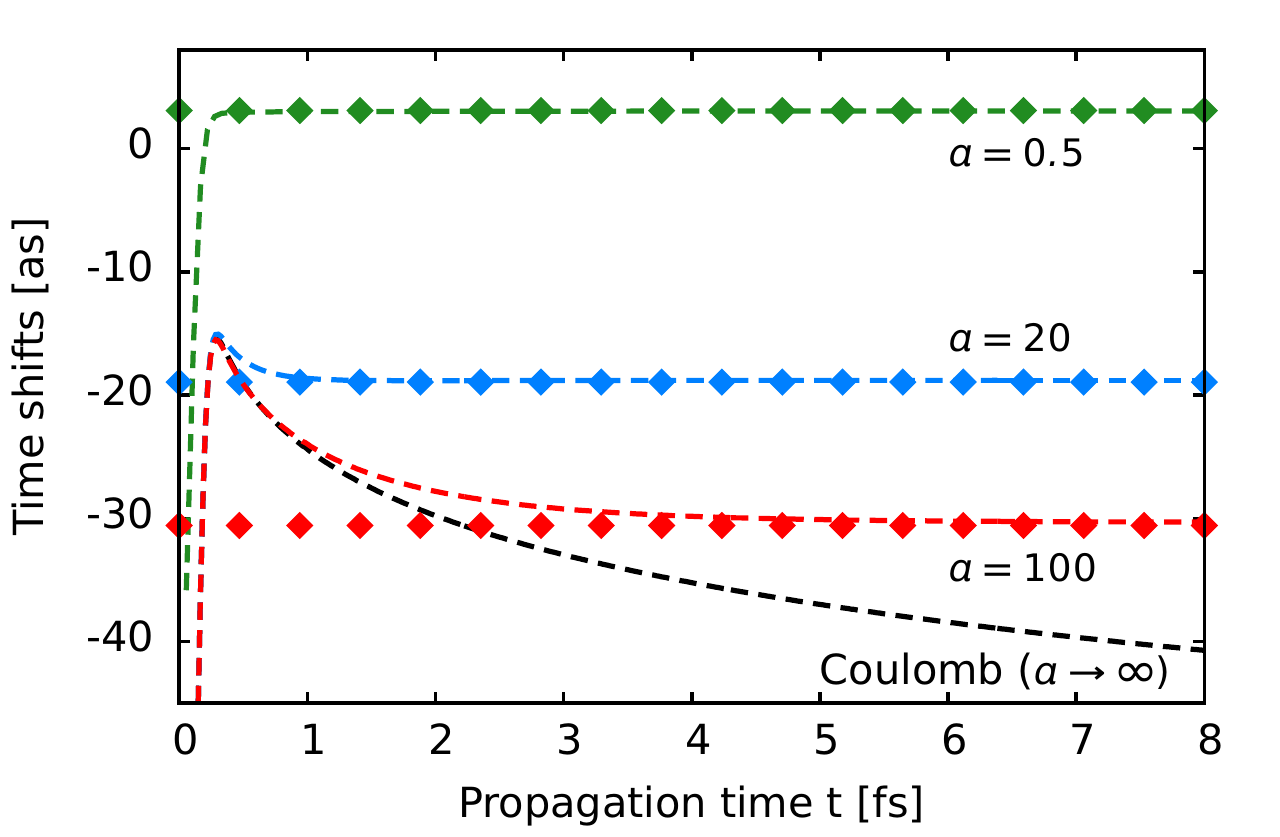}\label{fig:Yukawa_t_EWS}}
  \subfloat[][]{
  \includegraphics[width=0.5\linewidth]{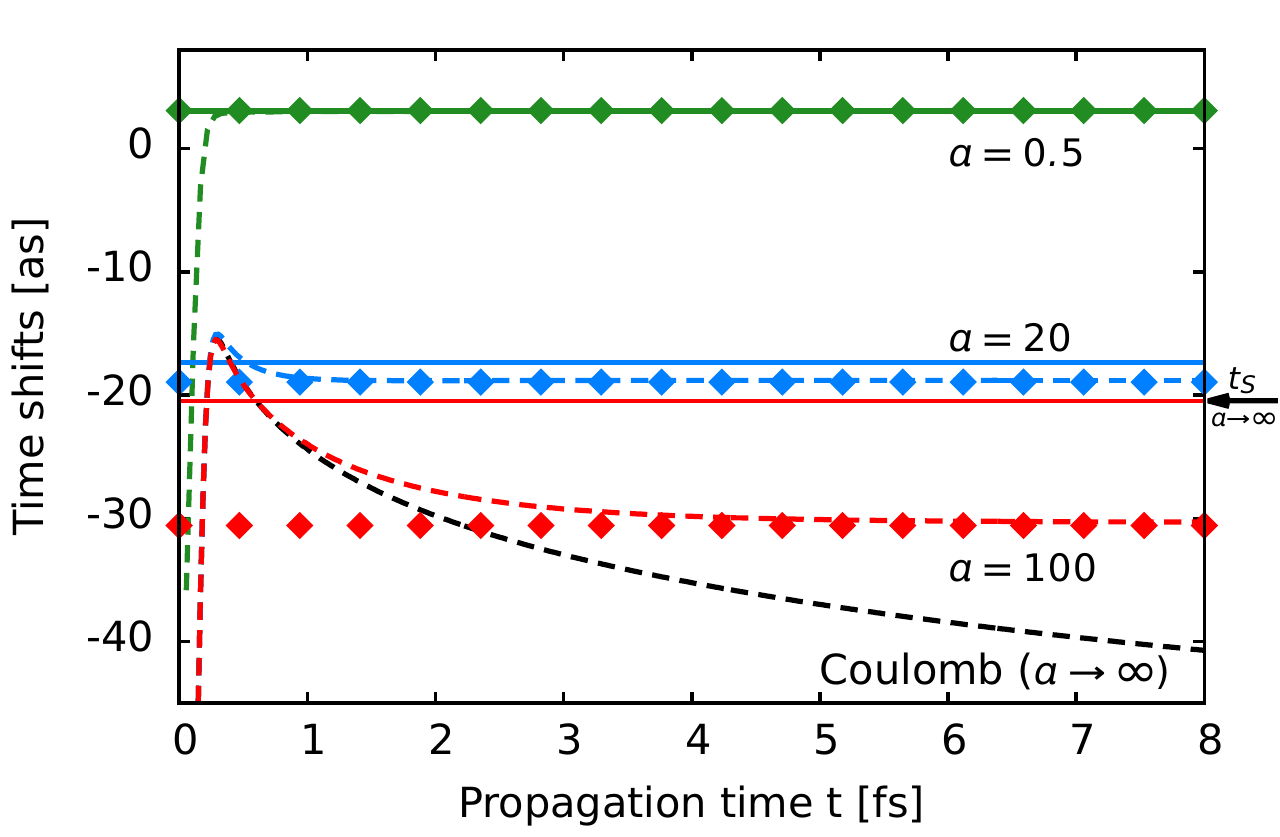} \label{fig:Yukawa_t_streaking}}
  \caption{Comparison of time shifts for ionization from a Coulomb potential with $Z=2$ and Yukawa potentials (\autoref{eq:Yukawa_potential})
with the same binding energy ($-54.4\ev$) but different screening lengths $a$. 
The ionizing XUV pulse has a duration of $200\as$, an intensity $I=10^{13}\Wcm$, and a central energy $\hbar\omega = 100 \eV$.\\
(a) The EWS time delay $\tEWS$ extracted from the
spatial wave packet (\autoref{eq:EWS_rkt}) created by the XUV pulse as a function of propagation time
𝑡(dashed lines) is compared to the EWS delay as extracted from the dipole matrix element \autoref{eq:EWS_def_matel} (\ie directly from the time-independent asymptotic scattering phase) which is spectrally averaged over the XUV photoelectron distribution around $45.6\ev$ (dots).
For the short-ranged Yukawa potential the two extraction methods for $\tEWS$ eventually agree, regardless of the value of $a$,  as the time $t$ when the wave packet is evaluated increases (compare the green ($a=0.5$), blue ($a=20$), and red ($a=100$) lines).
However, for the long-ranged Coulomb potential (black dashed line, $a\to\infty)$) a finite value for $\tEWS$ cannot be reached using \autoref{eq:EWS_rkt} (see text). \\
(b) Comparison of the EWS delays (dashed lines, dots) from (a) with the streaking time shifts $\tS$ (solid lines) for the same screening parameters. Only for small values of $a \lesssim 20$ (depending on the XUV energy) agreement is found (see also \autoref{fig:Yukawa_limit}). The streaking time shift for $a=100$ (red solid line) already coincides with the value in the Coulomb limit $a \to \infty$ (black arrow).  
The probing streaking IR field has a total duration of $6\fs$, a wavelength $\lambda = 800\nm$, and an intensity 𝐼$ I = 10^{12}\Wcm$. 
	\label{fig:Yukawa_t}}
\end{figure}

The additional logarithmic distortion of the wavefront (\autoref{eq:Coulphase_full}) gives rise to an additional time shift $\Delta \tCoul$ that depends on the radial coordinate $r$ or, likewise, on the time $t$ since $r \approx kt$,
\begin{equation}\label{eq:logarithmic_distortion}
\Delta \tCoul (E,r) = \frac{Z}{(2E)^{3/2}} \left[1 - \ln (2\sqrt{2E} r) \right] = \frac{Z}{(2E)^{3/2}} \left[1 - \ln (4 E t) \right]\, .
\end{equation}
Therefore, for Coulomb potentials EWS delays as obtained from the expectation value (crest position) of the radial wave packet (\autoref{eq:EWS_rkt}) do not converge to a finite value as $t \to \infty$ and always depend on the time of evaluation (\autoref{fig:Yukawa_t_EWS}).
However, it is exactly the time shift $\Delta \tCoul$ (\autoref{eq:logarithmic_distortion}) that gives rise to the $\tCLC$ contribution in streaking (\autoref{eq:delay_relation_full}). Key is that the streaking laser field maps the $\Delta\tCoul (E,r=kt)$ contribution 
onto the streaking time shift, however, only for a short time interval $t_0$ covering a small fraction of the IR laser field period $T_\mathrm{IR}$, $t_0 \ll T_\mathrm{IR}$, within which the IR field changes relatively little.
It has been numerically shown \cite{PazNagBur2013} that the CLC contribution for a wide range of $Z$ and $E$ is accurately given by
\begin{equation} \label{eq:tCLC_cl}
 \tCLC(E) \approx \Delta \tCoul (E,r=kt_0) = \frac{Z}{(2E)^{3/2}}\left(1 - \ln(0.37 E T\subscr{IR})\right) \, ,
\end{equation}
with $t_0 \approx 0.09 T_\mathrm{IR}$.
For short-ranged binding potentials as in the case of photodetachment the CLC contribution (\autoref{eq:tCLC_cl}) is absent.
In view of the fact that the CLC contribution to the streaking time is sensitive only to that portion of the atomic binding potential that is probed within the time interval $t_0$, it is of conceptual interest to probe the transition regime from short-ranged to long-ranged potentials.
We consider binding potentials that are still asymptotically short-ranged, however, mimicking the Coulomb potential over longer distances for which the traversal time of the emitted electron may become comparable to $t_0$.
We explore this connection in \autoref{sec:transition} for an exponentially screened Coulomb potential with varying screening length. \\

Turning to the residual correction terms in \autoref{eq:delay_relation_full}, 
the dipole-laser coupling contributions $\tdLC^{(i)}$ and $\tdLC^{(\mathrm{e-e)}}$ in \autoref{eq:delay_relation_full} become important when the initial or final ionic states are highly polarizable \cite{BagMad2010,BagMad2010Err}. For special cases, \ie initial or final states in degenerate hydrogenic manifolds, analytical expressions for their values can be given \cite{PazFeiNag2012,PazNagBur2013}. Clearly, the final state dipole-laser coupling term $\tdLC^{(\mathrm{e-e)}}$ is only present for multi-electron systems. 
One example of the latter type will be discussed below for photoemission from the central atom of an endohedral complex A@$\CC$ where the polarization of the buckyball strongly influences the streaking time.

\section{Transition to long-ranged potentials}\label{sec:transition}

\begin{figure}[tbh]
  \centering
  \includegraphics[width=0.6\linewidth]{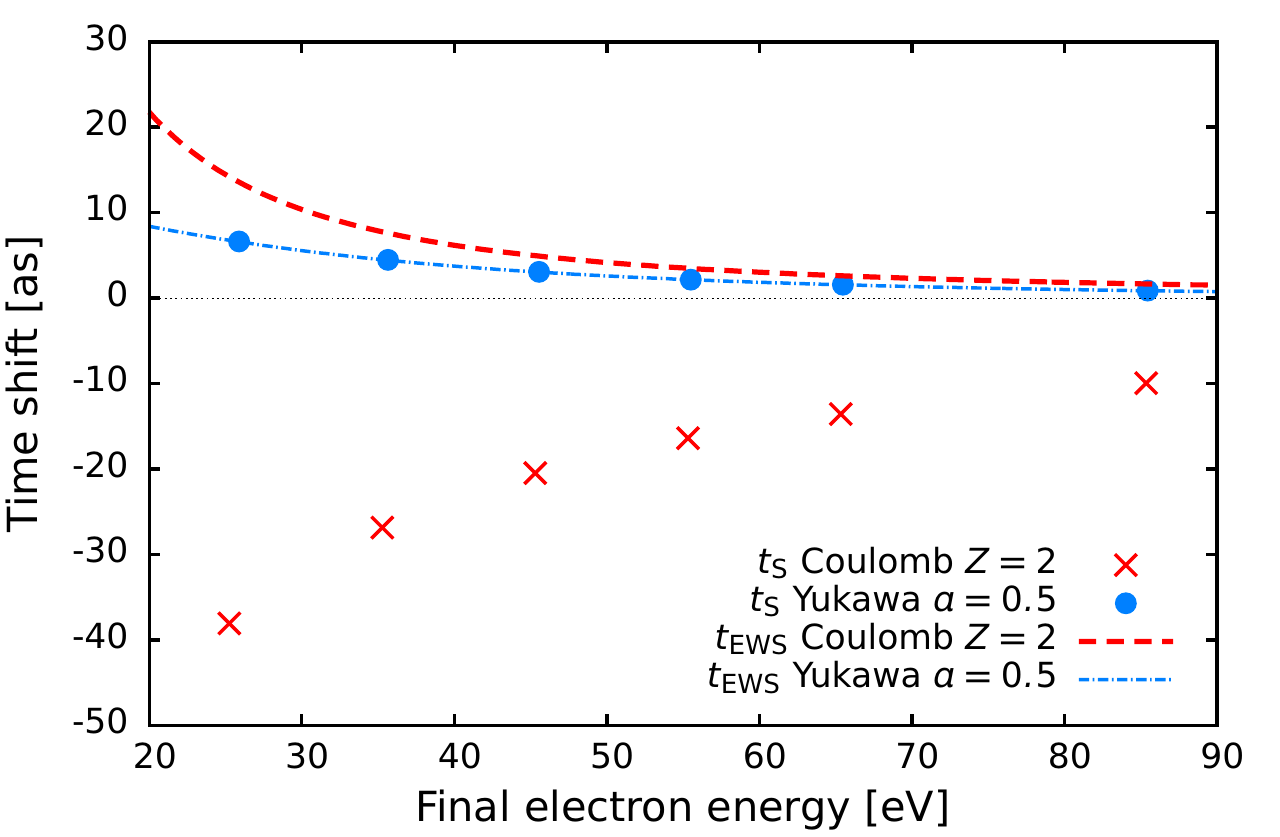}
  \caption{Streaking time shifts $\tS$ (\autoref{eq:streaking_fit}) and EWS delays $\tEWS$ (\autoref{eq:EWS_def_matel}, \autoref{eq:EWS_def_Coulphase_l}) for photoionization (or photodetachment) of an $1s$ initial state for different photon energies from a $\Hep$ Coulomb potential (red) with $Z=2$ and a Yukawa potential (blue) with the screening length $a=0.5$ (\autoref{eq:Yukawa_potential}). The streaking IR laser field has a wavelength of 800\,nm, a duration of 3\,fs, and an intensity of $10^{12}$\,W/cm$^2$.
	\label{fig:Yukawa-Coulomb-comparison}}
\end{figure}

\begin{figure}[htb]
  \centering
  \includegraphics[width=0.6\linewidth]{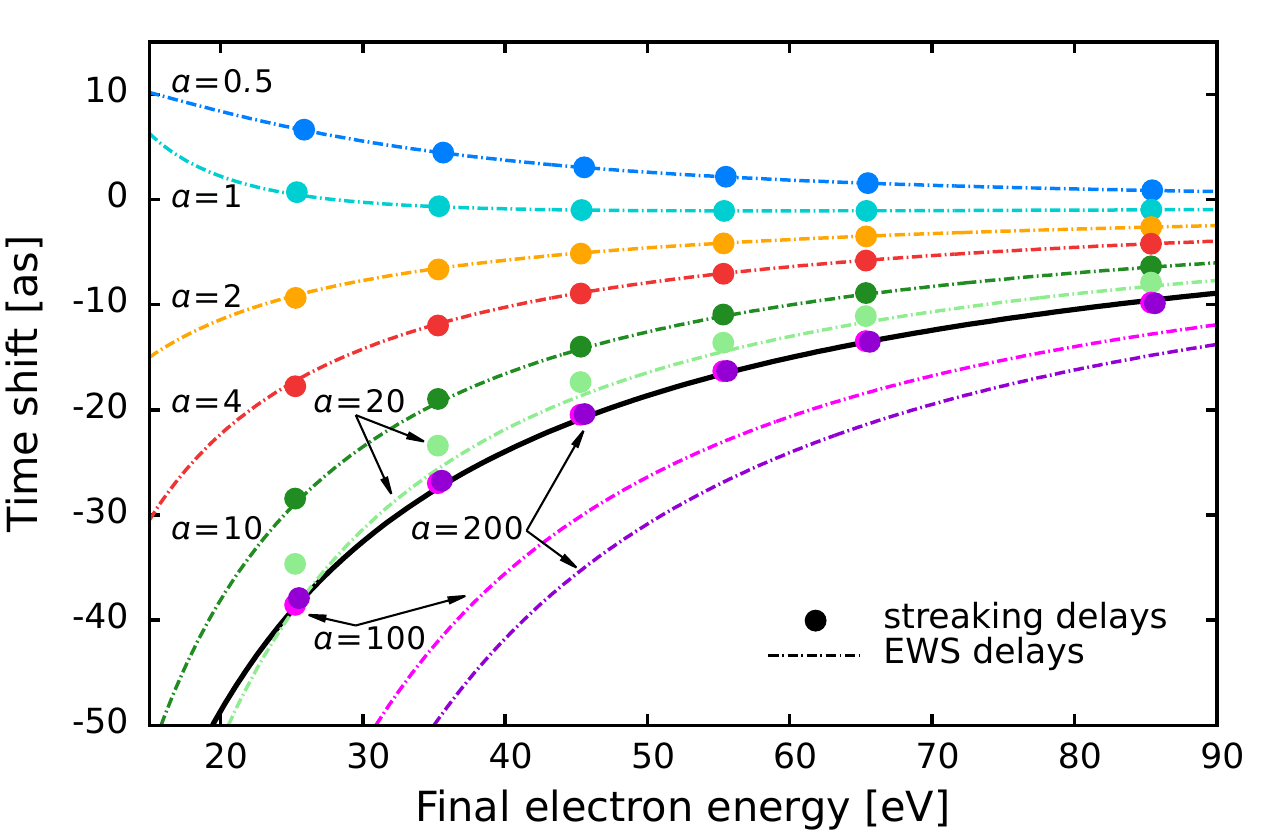}
  \caption{Streaking time shifts $t_S$ (dots, \autoref{eq:streaking_fit}) and EWS delays $\tEWS$ (dash-dotted lines, \autoref{eq:EWS_def_matel}) as a function of the final electron energy $E$ for short-ranged Yukawa potentials $V_\mathrm{Y}(r)$ (\autoref{eq:Yukawa_potential}) for various screening lengths approaching the limit $a\to \infty$. The charge $Z$ in \autoref{eq:Yukawa_potential} is adjusted to keep the binding energy constant at $-54.4\ev$. The streaking IR laser field has a wavelength of 800\,nm, a duration of 3\,fs, and an intensity of $10^{12}$\,W/cm$^2$.
The streaking time shifts $\tS$ approach the Coulomb limit (solid black line) for $a\to \infty$.
The agreement with the EWS time shifts $\tEWS$ (dash-dotted lines), which do not continuously reach an asymptotic limit, gradually decreases for $a\to \infty$. 
The EWS delays $\tEWS$ for the Yukawa potential do not converge to the Coulomb EWS delay (\autoref{eq:EWS_def_Coulphase_l}) in the limit $a\to \infty$ since the logarithmic $r$-dependent term (\autoref{eq:logarithmic_distortion}) due to the long-range potential tail is not included in the definition of the Coulomb EWS delay.\label{fig:Yukawa_limit}}
\end{figure}

In a previous contribution \cite{NagPazFei2011} we have shown that for photodetachment from short-ranged 
exponentially screened Coulomb (or Yukawa) potentials,
\begin{equation}
\label{eq:Yukawa_potential}
V_\mathrm{Y}(r) = - \frac{Z}{r} \exp(-r/a) \, ,
\end{equation}
with screening lengths $a$ of the order of a typical atomic radius,
the streaking time shifts $\tS$ equal the EWS time shifts $\tEWS$, and \autoref{eq:delay_relation_full} thus reduces,
in the absence of any dipole-laser coupling, to (see \autoref{fig:Yukawa-Coulomb-comparison})
\begin{equation} \label{eq:delay_relation_Yukawa}
  \tS = \tEWS  \, .
\end{equation} 
In contrast, for photoionization from long-ranged Coulomb potentials 
\begin{equation}
\label{eq:Coulomb_potential}
V_\mathrm{C}(r) = - \frac{Z}{r} \, ,
\end{equation}
the streaking time shifts $\tS$ and the EWS delays $\tEWS$ do not agree (\autoref{fig:Yukawa-Coulomb-comparison}) and the additional IR-field induced contribution, $\tCLC$ (\autoref{eq:tCLC_cl}), is present 
\begin{equation} \label{eq:delay_relation_Coulomb}
  \tS = \tEWS + \tCLC \, .
\end{equation}
As now $a$ is increased and $\tEWS$ becomes a function of $a$, one may expect a transition from the limit of \autoref{eq:delay_relation_Yukawa} to the long-range limit of \autoref{eq:delay_relation_Coulomb}.
To study this regime in more detail, we perform streaking simulations for several XUV energies and varying screening lengths $a$ (\autoref{fig:Yukawa_limit}). In all cases the charge $Z$ in \autoref{eq:Yukawa_potential} is adjusted to keep the ionization potential $I_p$ of the initial $1s$ state at the constant value $I_p = 54.4\eV$ to allow comparison with ionization from the ground state of the Coulombic $\Hep$ ion.
As the parameter $a$ gets larger, the curves for the EWS delays are shifted to larger negative delays (\autoref{fig:Yukawa_limit}). While for small $a$ excellent agreement between the EWS delays and the streaking time shifts persists (\autoref{eq:delay_relation_Yukawa}), noticeable deviations appear for larger $a$. Those discrepancies between $\tEWS$ and $\tS$ depend on the energy of the photoelectron where the lower energies are more heavily affected.
Finally, the streaking time shifts become insensitive to a further increase of the screening parameter $a$. At that point they converge to the streaking delays as obtained for the $\Hep$ Coulomb potential which corresponds to the $a \to \infty$ limit in \autoref{eq:Yukawa_potential}.
Figs.\ \ref{fig:Yukawa_t_streaking} and \ref{fig:Yukawa_limit} highlight the finite range of the atomic binding potential attosecond streaking is capable of probing.
The coupling to the probing IR field is sensitive to the binding potential close to the atomic core but not at asymptotic distances.
This is consistent with the estimate for $\tCLC$ (\autoref{eq:tCLC_cl}) \cite{PazNagBur2013}. 
The time interval $t_0$ during which the Coulomb-laser coupling is effective corresponds to a traversal time of the electron through the atomic force field, $t_0 \approx a/v$.
In turn, the range $a$ that can be probed ($a \lesssim 20\au$ for the results presented in \autoref{fig:Yukawa_limit}) depends on the electron velocity $v$ and $\lambda_\mathrm{IR}$ (since $t_0$ depends on $\lambda_\mathrm{IR}$, \cf \autoref{eq:tCLC_cl}).

\section{Attosecond streaking of photoemission from endohedral fullerenes}

Photoemission from the central atom of an endohedral buckyball A@$\CC$ represents an interesting and challenging case of attosecond streaking of a complex many-electron system.
The confinement resonances closely related to EXAFS (extended X-ray absorption fine structure) have been detected in the spectral domain \cite{KilAguMul2010}.
They are expected to leave their signature on the streaking spectrogram in the time domain as well.
However, in this complex system additional effects including final-state dipole laser coupling (\autoref{eq:delay_relation_full}), screening, and transport have to be accounted for.

\begin{figure}[htb]
  \centering
  \subfloat[][]{
  \includegraphics[width=0.5\linewidth]{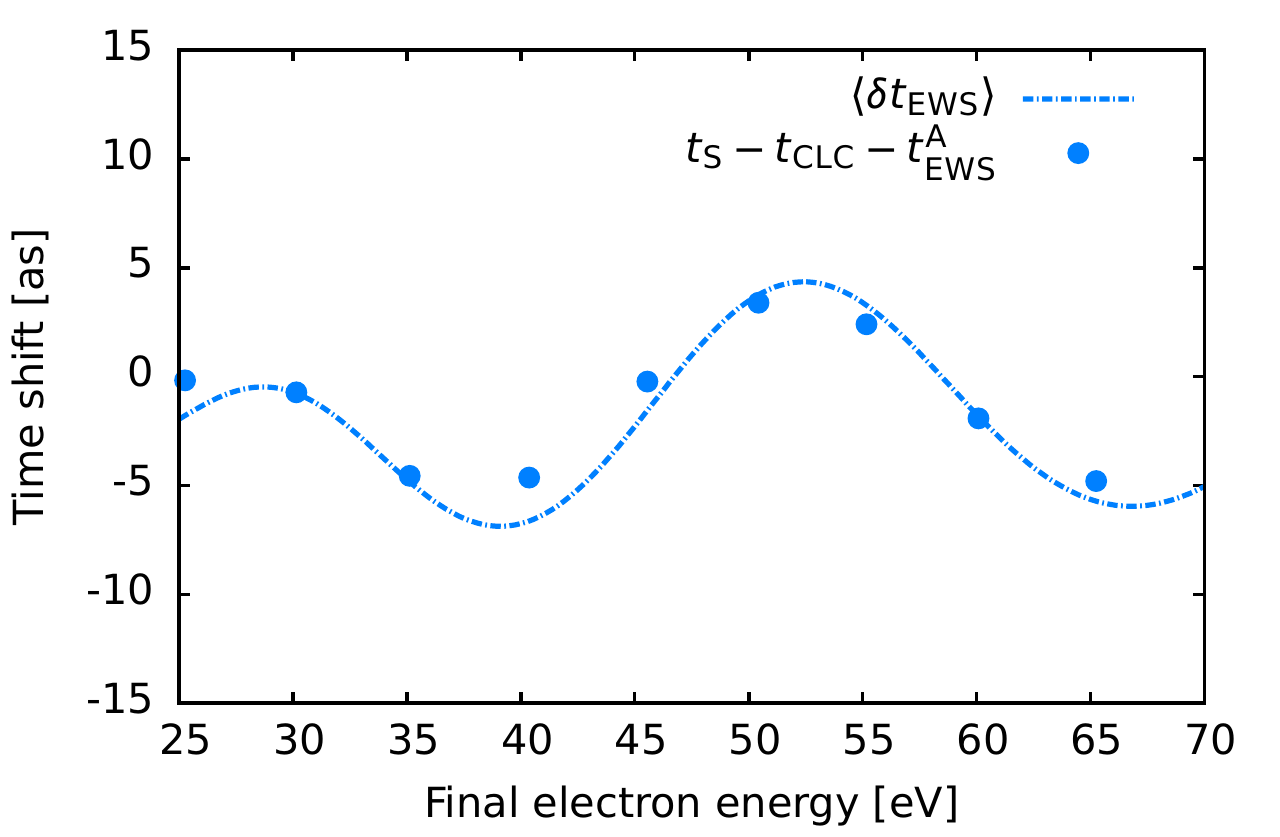}\label{fig:endohedral_EWS}}
  \subfloat[][]{
  \includegraphics[width=0.5\linewidth]{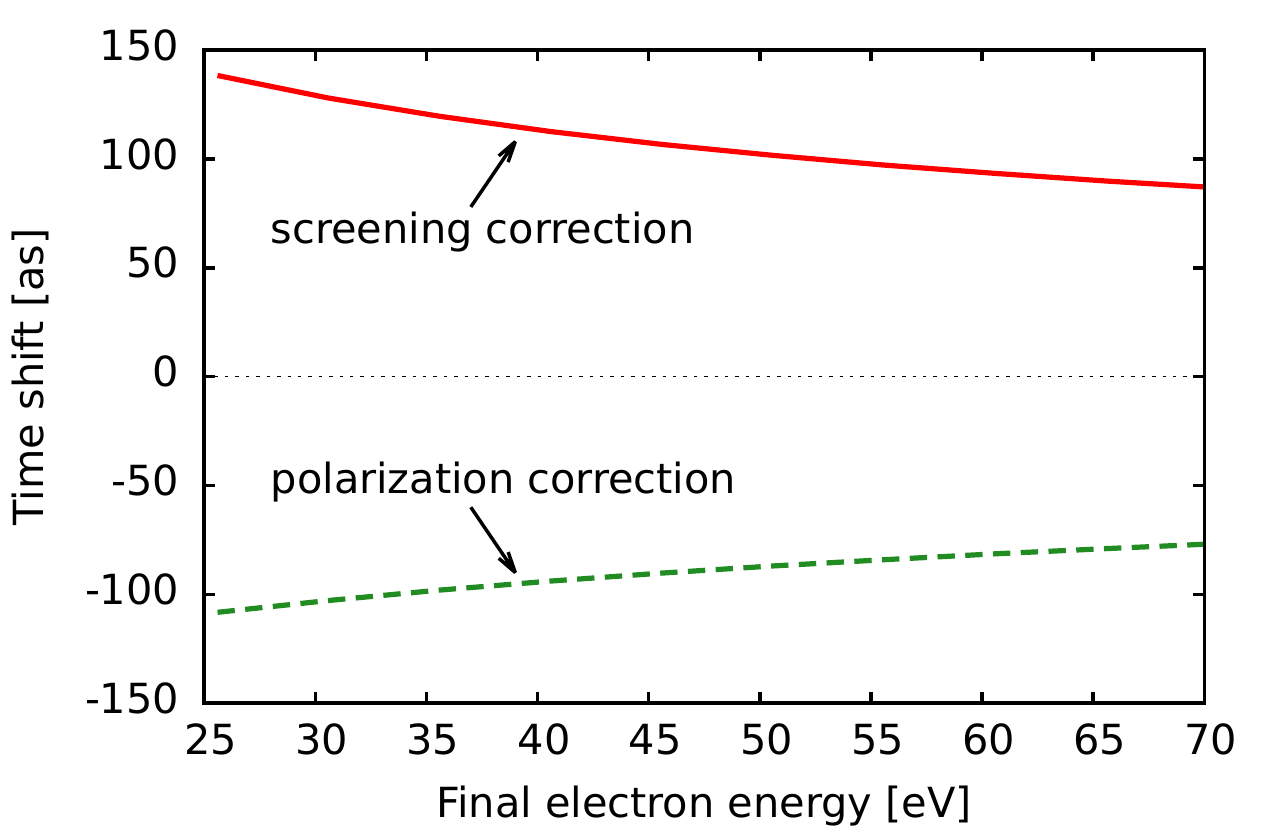} \label{fig:endohedral_corr}}
  \caption{Time resolved photoemission from endohedral fullerenes $\Hep$@$\CC$ (\cf the isolated $\Hep$ case in \autoref{fig:Yukawa-Coulomb-comparison}).
	(a) Modulation of $\tEWS$ (lines, averaged over the XUV spectrum) due to confinement resonances as a function of the kinetic energy of the emitted photoelectron (\autoref{eq:EWS_delta}) as extracted from the dipole matrix element (dash-dotted line) or from the streaking spectrograms (dots).
        For $V\subscr{shell}$ we have used values from the literature \cite{DolKinOgl2012}: $V_0=-0.302\au$, $\Delta=1.9\au$, $r_0=5.89\au$
	(b) Additional correction terms that contribute to the streaking time shift $\tS$ when the screening of the probing IR field inside the fullerene is accounted for (red, solid line) or the $\CC$ cage is polarized by the IR field giving rise to an additional dipole-laser coupling that the photoelectron encounters, $\tdLC$ (green, dashed line).\label{fig:endohedral}}
\end{figure}

We first model the $\CC$ cage by a spherically symmetric square-well potential $V\subscr{shell}(r)$ that the photoelectrons encounter after photoionization.
The model potential for the shell is given by (\cf \cite{DolKinOgl2012})
\begin{equation}
\label{eq:C60_pot}
V\subscr{shell}(r) = 
\begin{cases}
-V_0\quad \mathrm{for} \quad r_0 \le r \le r_0 + \Delta \\
0 \quad \mathrm{otherwise}
\end{cases}
\end{equation}
where $r_0$ is the inner-radius of the $\CC$ shell, $\Delta$ is its width, and $V_0$ the potential depth. 
The short-ranged admixture $V\subscr{shell}(r)$ (\autoref{eq:C60_pot}) to the Coulomb potential of the confined atom (\autoref{eq:Coulomb_potential}) leads to an energy dependent modulation $\delta\tEWS$ of the intrinsic atomic time delay $\tEWS^A$,
\begin{equation} \label{eq:EWS_delta}
\tEWS(E) = \tEWS^A(E) + \delta\tEWS(E)
\end{equation}
on the fully coherent one-electron level. The modulation $\delta\tEWS$ is the signature of the confinement resonances (\autoref{fig:endohedral_EWS}).
To facilitate the quantitative comparison with the atomic results (\autoref{fig:Yukawa_limit}) we study $\Hep$@$\CC$.
We note that this ionic rather than neutral central atom would not be stable.
The resulting streaking delays for the endohedral fullerenes nicely reflect the oscillations due to the confinement resonances (\autoref{fig:endohedral_EWS}) which implies that
\begin{equation}
\tS = \tEWS^A(E) + \delta\tEWS(E) + \tCLC
\end{equation}
is fulfilled to a high degree of accuracy for this model system.

A more realistic treatment of streaking for endohedral fullerenes includes the modification of the external IR field due to its interaction with the $\CC$ cage.
First, the streaking IR field will be considerably screened in the inside of the fullerene, and second, the $\CC$ shell is highly polarizable.
The first effect bears resemblance to streaking of photoelectrons from surfaces \cite{CavMueUph2007,LemSolTok2009} where electrons are effectively exposed to the streaking field only after traversing condensed matter and exiting the surface.
In the present case, the surface is provided by the fullerene shell.
To estimate the influence of the screening inside the $\CC$ cage on the streaking delays we perform classical trajectory calculations (\cf \cite{NagPazFei2011}) where the probing IR field is completely suppressed inside the cage. 
The resulting correction term (\autoref{fig:endohedral_corr}), \ie the difference of the streaking time shifts to those without screening is large and is of the order of $\approx +100\as$. 
Quantum mechanical simulations for a screened potential lead to virtually the same results (not shown).
The size of the screening correction is determined by the runtime the liberated photoelectron needs until reaching the streaking field outside the fullerene (\ie the apparent birth time in the continuum is delayed).
Moreover, the CLC term is modified since contributions from the logarithmic phase term (\autoref{eq:logarithmic_distortion}) are only acquired for $r > r_0$ (\cf \autoref{fig:Yukawa_t_EWS}). 

The IR-field induced polarization of the residual $\CC$ cage gives rise to a strong dipole-laser coupling the photoelectron encounters after leaving the buckyball.
The additional induced dipole potential outside the $\CC$ cage
\begin{equation}
\label{eq:dipole_potential}
\Phi(\cvec r, t) = \frac{\cvec p(t) \cdot \cvec r}{r^3}
\end{equation}
is determined by the dipole moment $\cvec p(t) = \alpha \cvec E_\mathrm{IR}(t)$ where $\alpha$ is the polarizability of $\CC$.
A classical trajectory simulation of streaking in the resulting spatially inhomogeneous dipole field 
yields a strong negative time shift, $\tdLC$, which is much larger than its atomic analogue (\autoref{fig:endohedral_corr}).
It is of the order of $\approx -100\as$.
Again, quantum-mechanical streaking simulations featuring an additional external dipole field lead to virtually the same delay times as the classical simulations.
Remarkably, screening and polarization effects are both large but of opposite sign partially canceling each other.
Consequently, attosecond streaking of photoionization of endohedral fullerenes $A$@$\CC$ does not only probe the EWS time shift (\autoref{fig:endohedral_EWS}) but is also sensitive to the run-time of the photoelectron in the screened $\CC$ cage and to the induced-dipole field of the IR-field polarized buckyball. 

\section{Summary}

We have reviewed time-resolved photoemission as probed by attosecond streaking and analyzed the relation between the measured or simulated streaking time $\tS$ and the intrinsic atomic Eisenbud-Wigner-Smith (EWS) time delay $\tEWS$.
Several IR-probe field induced contributions can be accounted for.
They include the Coulomb-laser coupling (CLC) and the dipole-laser coupling (dLC).
We have investigated how the CLC time shift is modified in the transition from short-ranged to long-ranged potentials.
We have found that attosecond streaking probes only the part of the binding potential close to the nucleus. 
Furthermore, we could show that the confinement resonances in the ionization of endohedral fullerenes can be probed in the time domain using attosecond streaking.
We identified corrections due to
screening of the probing infrared field inside the $\CC$ cage and the dynamical polarization of $\CC$ by the streaking field. 
Further theoretical studies of this scenario with the help of density-functional theory are under way aiming for the non-linear response of the buckyball to the IR field and a more realistic model for the $\CC$ scattering potential. \\

The authors would like to thank Johannes Feist, Katharina Doblhoff-Dier, Christoph Lemell, and K\'{a}roly T\H{o}k\'{e}si for valuable input. 
This work was supported by the FWF-Austria (SFB NEXTLITE, SFB VICOM, and P23359-N16), and in part by the National Science Foundation through XSEDE resources provided by NICS and TACC under Grant TG-PHY090031. The computational results presented have also been achieved in part using the Vienna Scientific Cluster (VSC).
\pagebreak

\providecommand{\newblock}{}

\end{document}